\documentclass[aps, prb, twocolumn, amssymb, amsmath, showpacs, superscriptaddress]{revtex4-1}
\usepackage{bm}
\usepackage{times}
\usepackage{graphicx}
\usepackage{color}
\usepackage{dcolumn}
\usepackage[colorlinks=true, letterpaper=true, pdfstartview=FitV, linkcolor=blue, citecolor=blue, urlcolor=blue]{hyperref}
\usepackage{appendix}
\usepackage[normalem]{ulem}

\newcommand{\vect}[1]{\boldsymbol{#1}}                 
\usepackage{helvet}
\usepackage{pifont}
\newcommand{\cmark}{\ding{51}}
\newcommand{\xmark}{\ding{55}}
\usepackage{multirow}
\usepackage{siunitx}
\usepackage{makecell}

\usepackage{colortbl}
\usepackage{xcolor}
\definecolor{lightgreen}{rgb}{0.88, 1.0, 0.88}

\begin{document}

\title{Quantum geometric map of magnetotransport}
\author{Longjun Xiang}
\affiliation{College of Physics and Optoelectronic Engineering, Shenzhen University, Shenzhen 518060, China}
\author{Jinxiong Jia}
\affiliation{Department of Physics, University of Science and Technology of China, Hefei, Anhui 230026, China}
\author{Fuming Xu}
\email{xufuming@szu.edu.cn}
\affiliation{College of Physics and Optoelectronic Engineering, Shenzhen University, Shenzhen 518060, China}
\affiliation{Quantum Science Center of Guangdong-Hongkong-Macao Greater Bay Area (Guangdong), Shenzhen 518045, China}
\author{Jian Wang}
\email{jianwang@hku.hk}
\affiliation{College of Physics and Optoelectronic Engineering, Shenzhen University, Shenzhen 518060, China}
\affiliation{Quantum Science Center of Guangdong-Hongkong-Macao Greater Bay Area (Guangdong), Shenzhen 518045, China}
\affiliation{Department of Physics, The University of Hong Kong, Pokfulam Road, Hong Kong, China}


\begin{abstract}
We propose a quantum geometric map for the magnetononlinear Hall effect (MNHE), the planar Hall effect (PHE),
and the ordinary Hall effect (OHE).
These magnetotransport phenomena originate from the bilinear charge current of Bloch electrons in electromagnetic fields,
incorporating both spin Zeeman coupling and orbital minimal coupling to the applied magnetic field.
Benchmarked against Onsager reciprocity, we demonstrate that the spin- and orbital-induced MNHEs
are governed by the time-reversal-even Zeeman quantum metric dipole and conventional quantum metric quadrupole, respectively;
the spin- and orbital-induced PHEs are dominated by the time-reversal-odd
Zeeman Berry curvature dipole and conventional Berry curvature quadrupole, respectively.
We further show that the OHE contains an interband contribution that is
related to the quantum metric quadrupole, contrary to conventional wisdom.
Navigated by this map, we study the previously unexplored
spin-induced PHE in the surface Dirac cone of topological insulators,
where we uncover a step-like PHE.
Our work offers a unified quantum geometric framework for understanding magnetotransport experiments.
\end{abstract}

\maketitle

\begin{figure}[t!]
\includegraphics[width=0.95\columnwidth]{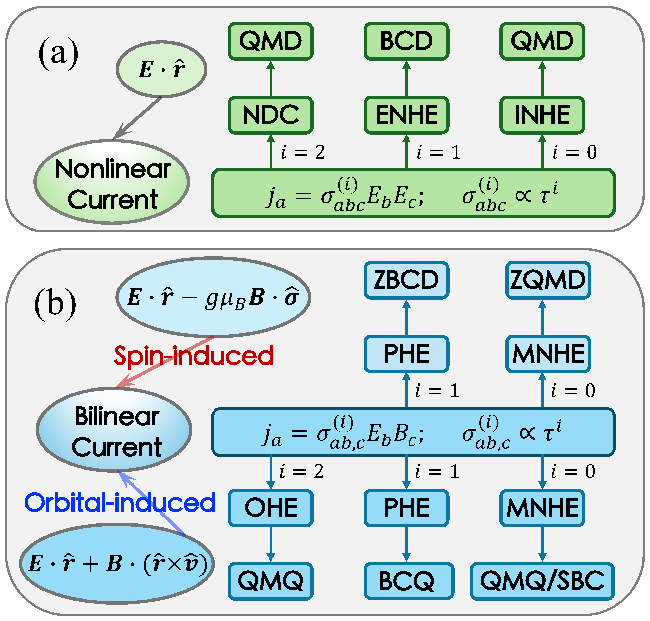}
\caption{
Quantum geometric map of (a) the nonlinear charge current Eq.~(\ref{EEresponse})
and (b) the bilinear charge current Eq.~(\ref{EBresponse}).
NDC~\cite{FuBCD}: nonlinear Drude current;
ENHE~\cite{FuBCD}/INHE~\cite{GaoY2014PRL}: extrinsic/intrinsic nonlinear Hall effect;
PHE~\cite{PHE, MaDaPRB, OrtixPHE, YaoPHE}: planar Hall effect;
MNHE~\cite{GaoY2014PRL, MNHE1, MNHE2, MNHE3, MNHE4, MNHE5}: magnetononlinear Hall effect;
OHE~\cite{Nagaosa2010}: ordinary Hall effect;
BCD~\cite{FuBCD}: Berry curvature dipole;
ZBCD~\cite{xiangSHE}; Zeeman BCD; 
QMD~\cite{BPT1}: quantum metric dipole;
ZQMD~\cite{xiangSHE}: Zeeman QMD;
BCQ~\cite{BCQ1, BCQ2, BCQ3}: Berry curvature quadrupole;
QMQ~\cite{QMQ1,QMQ2,QMQ3}: quantum metric quadrupole;
SBC~\cite{Duan2024}: square Berry curvature.
}
\label{FIG1}
\end{figure}

\noindent \textit{\textcolor{blue}{Introduction}}---The current responses of Bloch electrons
in crystalline solids under electromagnetic fields
are profoundly entrenched in quantum geometry~\cite{Xiao2010, QuantumGeometry, YanPRR, QuantumGeometry1, QuantumGeometry2, MaQlight,
JTsong, AMIT, XiaoCQF, Rashbagas, entropy, PRA, Berniverg, YanReview, HZLu2024, Fontana, SuYangXuRMP, DaoDao, QueirozReview}.
\textcolor{red}{Within the non-interacting single-particle framework and the relaxation time ($\tau$) approximation,
the second-order nonlinear charge current $j_a$ in a DC electric field $E_b$
can be expressed as}~\cite{xiangSHE, GaoYFOP}
\begin{align}
j_a=\sigma_{abc}^{(i)}E_bE_c,
\label{EEresponse}
\end{align}
where $i=0, 1, 2$ labels the order in $\tau$ of the conductivity tensor~\cite{tau},
i.e., $\sigma_{abc}^{(i)} \propto \tau^i$.
The term $\sigma_{abc}^{(0)}$, induced by the $\mathcal{T}$-odd (time-reversal-odd) quantum metric dipole (QMD),
leads to the intrinsic nonlinear Hall effect (INHE)~\cite{GaoY2014PRL, BPT1, BPT2,
XuSY2023, Wang2023, Han2024, Wang2023, YanPRL2024, CulcerPRBL, GangSu2022, Jia2024},
whereas the term $\sigma_{abc}^{(1)}$,
induced by the $\mathcal{T}$-even Berry curvature dipole (BCD),
causes the extrinsic nonlinear Hall effect (ENHE)~\cite{FuBCD, BCDexp1, BCDexp2, BCDexp3}.
Moreover, the interband contribution to the nonlinear Drude current (NDC),
described by $\sigma_{abc}^{(2)}$,
is also related to the QMD~\cite{QMDfootnote},
analogous to the interband linear Drude current arising from the quantum metric~\cite{NagaosaDrude}.
As a result, a quantum geometric map that captures the symmetry constraints~\cite{Psym} of these effects was developed,
as illustrated in Fig.~\ref{FIG1}(a).

In the same spirit, the bilinear charge current driven by both DC electric and magnetic fields
can be written as
\begin{align}
j_a=\sigma_{ab,c}^{(i)}E_bB_c,
\quad
\sigma_{ab,c}^{(i)} \propto \tau^i,
\quad
i=0, 1, 2.
\label{EBresponse}
\end{align}
Here the magnetic field $B_c$ can enter via orbital minimal coupling,
$\vect{B}\cdot(\hat{\vect{r}}\times\hat{\vect{v}})$,
and via spin Zeeman coupling, $\vect{B}\cdot\hat{\vect{\sigma}}$.
For orbital minimal coupling, $\sigma_{ab,c}^{(0)}$ and $\sigma_{ab,c}^{(1)}$ give rise to
the magnetononlinear Hall effect (MNHE)~\cite{GaoY2014PRL} and planar Hall effect (PHE)~\cite{PHE, MaDaPRB, OrtixPHE, YaoPHE}, respectively;
the former has recently been observed in a kagome magnet, Fe$_3$Sn$_2$~\cite{MNHE1}.
While the orbital-induced MNHE (PHE) is attributed to the quantum metric~\cite{MNHE1, MNHE2, MNHE3, MNHE4, MNHE5}
(Berry curvature~\cite{PHE, MaDaPRB, OrtixPHE, YaoPHE}),
a unified quantum geometric map that encodes their symmetry constraints is still lacking.
Further, the MNHE and PHE from spin Zeeman coupling are largely unexplored~\cite{MNHE1, MNHE2, MNHE3},
particularly in terms of quantum geometry.
Additionally, the interband quantum geometric contribution of $\sigma^{(2)}_{ab,c}$,
which is proportional to $\tau^2$ like the NDC and
corresponds to the familiar ordinary Hall effect (OHE)~\cite{Nagaosa2010}, has not been discussed.

In this work, benchmarked against Onsager reciprocity,
we propose a unified quantum geometric map [Fig.~\ref{FIG1}(b)] for these magnetotransport effects
on top of the Zeeman quantum geometry~\cite{xiangZeeman,Zeeman1, Zeeman2, Zeeman3, Zeeman4} and the conventional quantum geometry.
In particular, we show that the spin- and orbital-induced MNHEs
from the intrinsic bilinear current are governed by the $\mathcal{T}$-even Zeeman QMD (ZQMD)~\cite{xiangSHE} 
and conventional quantum metric quadrupole (QMQ)~\cite{QMQ1, QMQ2, QMQ3}, respectively.
In a quantum geometric dual way, the spin- and orbital-induced PHEs from the extrinsic (linear in $\tau$) bilinear current
are dominated by the $\mathcal{T}$-odd Zeeman BCD (ZBCD)~\cite{xiangSHE}
and conventional Berry curvature quadrupole (BCQ)~\cite{BCQ1, BCQ2, BCQ3}, respectively.
Furthermore, we find that the OHE (quadratic in $\tau$) from the Lorentz force indeed contains an interband contribution,
which arises from the conventional QMQ~\cite{QMQ1, QMQ2, QMQ3}
and can even dominate over the OHE observed in the two-dimensional Rashba electron gas.

After establishing the quantum geometric map for Eq.~(\ref{EBresponse}),
especially after carefully considering their symmetry constraints,
we study the previously unexplored spin-induced PHE using
the surface Dirac cone of three-dimensional topological insulators (TIs),
where its orbital counterpart is suppressed under an in-plane magnetic field.
Notably, the spin-induced PHE in the surface Dirac cone of TIs
features a step-like dependence on the chemical potential and
yields a large Hall voltage, offering a fingerprint to identify this effect.
Finally, possible applications based on this map are discussed.
Our work establishes quantum geometry as a universal framework for unifying magnetotransport
measurements in quantum materials.

\bigskip
\noindent{\textit{\textcolor{blue}{Onsager reciprocity of magnetotransport}}}---Before turning
to the quantum geometric origin of bilinear magnetotransport and further constructing its quantum geometric map,
we first discuss the general constraints on bilinear magnetotransport imposed by Onsager reciprocity~\cite{Onsager1, Onsager2, Onsager3}.
For the linear charge current response defined by $j_a = \sigma_{ab} E_b$,
the Onsager reciprocal relation reads~\cite{Onsager3}
\begin{align}
\sigma_{ab}(\vect{M}, \vect{B}) = \sigma_{ba} (-\vect{M}, -\vect{B}),
\label{onsager}
\end{align}
where $\vect{M}$ and $\vect{B}$ represent the internal magnetic order parameter (e.g., spontaneous magnetization)
and the external applied magnetic field, respectively. Without loss of generality,
we decompose $\sigma_{ab}$ into a symmetric component (denoted by $\sigma_{ab}^S$) 
and an antisymmetric component (denoted by $\sigma_{ab}^A$), as explicitly given by
\begin{align}
\sigma_{ab}^S \equiv \dfrac{\sigma_{ab}+\sigma_{ba}}{2},
\quad
\sigma_{ab}^A \equiv \dfrac{\sigma_{ab}-\sigma_{ba}}{2}.
\end{align}
Here $\sigma_{ab}^A$ yields the dissipationless Hall current~\cite{Nagaosa2010},
while $\sigma_{ab}^S$ governs the Joule heating since
$\vect{j}\cdot\vect{E}=j_aE_a=\sigma_{ab}E_aE_b=\sigma_{ab}^SE_aE_b$.
Combining this decomposition with Eq.~\eqref{onsager}, we immediately obtain
\begin{align}
\sigma_{ab}^S(\vect{M},\vect{B}) &= +\sigma_{ab}^S(-\vect{M},-\vect{B}), \label{sym} \\
\sigma_{ab}^A(\vect{M},\vect{B}) &= -\sigma_{ab}^A(-\vect{M},-\vect{B}), \label{asym}
\end{align}
which means that $\sigma_{ab}^S (\sigma_{ab}^A)$ is even (odd) under the combined reversal
$(\vect{M}, \vect{B}) \rightarrow (-\vect{M}, -\vect{B})$,
namely the $\mathcal{T}$-operation.

Constrained by Onsager reciprocity [Eqs.~\eqref{sym}-\eqref{asym}],
we conclude that only the symmetric component~\cite{anticomponent} $\sigma_{ab}^S$ is allowed
in a $\mathcal{T}$-invariant system ($\vect{M}=0$) at zero magnetic field ($\vect{B}=0$).
When a weak magnetic field $\vect{B}$ is applied, $\sigma^{A}_{ab}(\vect{B})$ becomes finite
and satisfies $\sigma_{ab}^A(\vect{B})=-\sigma_{ab}^A(-\vect{B})$.
Expanding $\sigma_{ab}^A(\vect{B})=\sigma_{ab,c}^AB_c+\mathcal{O}(|\vect{B}|^3)$,
where $\sigma_{ab,c}^A \equiv \partial \sigma_{ab}^A(\vect{B})/\partial B_c|_{\vect{B}=0}$,
we obtain the bilinear response equation $j_a=\sigma_{ab,c}^AE_bB_c$,
which yields the MNHE and OHE described by Eq.~\eqref{EBresponse}, as will be clear below.
From Eq.~\eqref{asym}, we find that $\sigma_{ab,c}^A$ is $\mathcal{T}$-even
and thus both MNHE and OHE can occur in $\mathcal{T}$-invariant systems.
Note that both MNHE and OHE are dissipationless since $\sigma_{ab,c}^A=-\sigma_{ba,c}^A$.
We stress that the MNHE is an intrinsic (independent of $\tau$) effect while
the OHE is an extrinsic (quadratically dependent on $\tau$) effect.

For a $\mathcal{T}$-broken system ($\vect{M} \neq 0$) at $\vect{B}=0$,
Eq.~\eqref{asym} reduces to $\sigma_{ab}^A(\vect{M})=-\sigma_{ab}^A(-\vect{M})$,
responsible for the anomalous Hall effect in ferromagnetic metals~\cite{Nagaosa2010}.
When both $\vect{M}$ and $\vect{B}$ are present,
expanding $\sigma_{ab}^S(\vect{M}, \vect{B})=\sigma_{ab}^S(\vect{M}, 0)+\sigma_{ab,c}^S(\vect{M})B_c+\mathcal{O}(|\vect{B}|^2)$,
where $\sigma_{ab,c}^S(\vect{M}) \equiv \partial \sigma_{ab}^S(\vect{M}, \vect{B})/\partial B_c|_{\vect{B}=0}$,
we obtain the bilinear response equation $j_a=\sigma_{ab,c}^S(\vect{M})E_bB_c$.
This also allows a transverse $(a \neq b)$ yet dissipative bilinear current in coplanar electromagnetic fields
and hence yields the PHE in Eq.~\eqref{EBresponse}, as will be clear below.
Using Eq.~\eqref{sym}, we find that $\sigma_{ab,c}^S(\vect{M})$ is $\mathcal{T}$-odd: $\sigma_{ab,c}^S(\vect{M})=-\sigma_{ab,c}^S(-\vect{M})$,
which dictates that the PHE can only appear in $\mathcal{T}$-broken systems.
Benchmarked against Onsager reciprocity,
we are ready to derive the quantum geometric expressions for MNHE, PHE, and OHE.

\bigskip
\noindent{\textit{\textcolor{blue}{Quantum geometric map of magnetotransport}}}---We
begin with the spin contribution to Eq.~\eqref{EBresponse}.
In the density matrix formalism~\cite{Sipe0, Sipe1}, the charge current is $j_a = \text{Tr}[\hat{\rho}\hat{v}^a]$,
where $\hat{v}^a$ is the velocity operator and $\hat{\rho}$ is the density matrix
obeying the quantum Liouville equation $i\partial_t \hat{\rho} = [H, \hat{\rho}]$
with $H=H_0+H_1$. Here $H_0$ is the Hamiltonian of the crystalline solid,
while $H_1=-g\mu_B\vect{B}\cdot\hat{\vect{\sigma}}+\vect{E}\cdot\hat{\vect{r}}$
is the perturbation due to external electromagnetic fields,
where $g$ is the $g$-factor, $\mu_B$ is the Bohr magneton,
$\hat{\vect{\sigma}}=(\hat{\sigma}^x, \hat{\sigma}^y, \hat{\sigma}^z)$ is the spin operator,
and $\hat{\vect{r}}$ is the position operator.
Throughout this work, we set $e=\hbar=1$ unless stated otherwise.

By iteratively solving the quantum Liouville equation for the density matrix elements in the Bloch basis of $H_0$,
defined by $H_0|u_n\rangle=\epsilon_n|u_n\rangle$,
where $\epsilon_n$ and $|u_n\rangle$ represent the band energy and the periodic Bloch state, respectively,
the bilinear charge current can be evaluated straightforwardly,
as detailed in the Supplementary Material~\cite{sup}.
By this calculation, the response tensors in Eq.~\eqref{EBresponse} are found to be
\begin{align}
\sigma^{(0)}_{ab,c}
&=
2g\mu_B
\sum_{nm} \int_k
\left(
\frac{v_n^a\mathcal{Q}_{nm}^{bc}}{\epsilon_{nm}} 
-
\frac{v_n^b\mathcal{Q}_{nm}^{ac}}{\epsilon_{nm}}
\right) 
f_n',
\label{spin0}
\\
\sigma_{ab,c}^{(1)}
&=
\frac{\tau}{2}g\mu_B
\sum_{nm} \int_k \left( v_n^a \mathcal{Z}_{nm}^{bc} + v_n^b \mathcal{Z}_{nm}^{ac} \right) f_n',
\label{spin1}
\end{align}
where $\int_k=\int d^dk/(2\pi)^d$ with $d$ being the spatial dimension,
$\epsilon_{nm}=\epsilon_n-\epsilon_m$,
$f_n' \equiv \partial f_n/\partial \epsilon_n$ with $f_n$ being the equilibrium Fermi-Dirac distribution function,
and $v_n^a=v_{nn}^a$ is the intraband velocity matrix element.
Additionally, $\mathcal{Q}^{bc}_{nl} \equiv \text{Re}[r^b_{nl}\sigma^c_{ln}]$ is the (local) Zeeman quantum metric~\cite{xiangZeeman},
whereas $\mathcal{Z}^{ac}_{nm}\equiv-2\text{Im}\left[ r^a_{nm} \sigma^c_{mn} \right]$ is the (local) Zeeman Berry curvature~\cite{xiangZeeman}.
Here $r^b_{nl}=\langle u_n|i\partial_b|u_l\rangle$ with $\partial_b=\partial/\partial k_b$
is the interband ($n \neq l$) Berry connection,
and $\sigma^c_{nl}=\langle n|\hat{\sigma}^c|l\rangle$ is the interband matrix element of spin operator $\hat{\sigma}^c$.

Eq.~\eqref{spin0} is antisymmetric under $a \leftrightarrow b$
and yields the spin-induced MNHE~\cite{MNHE1},
while Eq.~\eqref{spin1} is symmetric under $a \leftrightarrow b$
and produces the PHE when~\cite{BMRfoot} $a \neq b$ in coplanar electromagnetic fields.
By analogy with the INHE and ENHE induced by the QMD~\cite{BPT1} $v_n^a g^{bc}_{nm}$
and BCD~\cite{FuBCD} $v_n^a \Omega_{nm}^{bc}$, respectively,
we identify the quantum geometric origins of the spin-induced MNHE and PHE
as the ZQMD~\cite{xiangSHE} $v_n^a \mathcal{Q}^{bc}_{nm}$
and the ZBCD~\cite{xiangSHE} $v_n^a\mathcal{Z}_{nm}^{ab}$,
as summarized in the upper panel of Fig.~\ref{FIG1}(b).
Here $g^{bc}_{nm}=\text{Re}[r^{b}_{nm}r^c_{mn}]$ is the (local) quantum metric,
whereas $\Omega^{bc}_{nm}=-2\text{Im}[r^b_{nm}r^c_{mn}]$ is the (local) Berry curvature.
Notably, using~\cite{xiangZeeman} $\mathcal{T}\mathcal{Z}^{ab}_{nm}=\mathcal{Z}^{ab}_{nm}$,
$\mathcal{T}\mathcal{Q}^{ab}_{nm}=-\mathcal{Q}^{ab}_{nm}$, and $\mathcal{T}v_n^c=-v_n^c$,
we find that the ZQMD for the spin-induced MNHE is $\mathcal{T}$-even
while the ZBCD for the spin-induced PHE is $\mathcal{T}$-odd,
consistent with Onsager reciprocity discussed above.

For orbital minimal coupling, we show that the orbital-induced MNHE
is governed by the conventional QMQ~\cite{QMQ1, QMQ2, QMQ3}
$\epsilon_{aij}v_n^bv_n^ig^{jc}_{nm}$ (or $\epsilon_{aij}\partial^2_{bi}g^{jc}_{nm}$)
and by the square Berry curvature (SBC)~\cite{Duan2024} $\Omega^{ab}_{nm}\epsilon_{cij}\Omega^{ij}_{nm}$.
In contrast, the orbital-induced PHE is dominated by the conventional BCQ~\cite{BCQ1, BCQ2, BCQ3}
$v_n^av_n^b\epsilon_{cij}\Omega^{ij}_{nm}$ (or $\epsilon_{cij}\partial^2_{ab}\Omega^{ij}_{nm}$), 
as detailed in the Supplementary Material~\cite{sup} and
summarized in the lower panel of Fig.~\ref{FIG1}(b).
Here $\epsilon_{abc}$ is the Levi-Civita symbol, ensuring that
$\sigma_{ab,c}^{(i)}$ transforms as a rank-3 pseudotensor after performing the Einstein summation convention.
Note that the conventional QMQ and SBC for the orbital-induced MNHE are $\mathcal{T}$-even
while the conventional BCQ for the orbital-induced PHE is $\mathcal{T}$-odd,
using $\mathcal{T}g^{ab}_{nm}=g^{ab}_{nm}$ and $\mathcal{T}\Omega^{ab}_{nm}=-\Omega^{ab}_{nm}$,
also consistent with Onsager reciprocity.

In addition to MNHE and PHE, orbital minimal coupling can further generate the OHE
with the response tensor~\cite{sup}
\begin{align}
\sigma_{ab,c}^{(2)}
=
\tau^2 \sum_n \int_k \epsilon_{idc} v_n^a v_n^d \partial_i v_n^b f_n',
\label{orbital20}
\end{align}
which arises from the classical Lorentz force~\cite{Nagaosa2010} and has no spin counterpart.
Eq.~\eqref{orbital20} for the OHE is also consistent with the Onsager reciprocity
by noting that~\cite{sup} $\sigma_{ab,c}^{(2)}=-\sigma_{ba,c}^{(2)}$.
However, this conductivity has long been believed to be nongeometric~\cite{Duan2024}, similar to the Drude currents.
Using~\cite{QMDfootnote} $\partial_b v_n^a=v_{n}^{ab}+2\sum_m \epsilon_{nm} g^{ab}_{nm}$
with $v_n^{ab}=\langle u_n|\partial^2_{ab}H_0|u_n\rangle$,
Eq.~\eqref{orbital20} can be rewritten as
\begin{align*}
\sigma_{ab,c}^{(2)}
=
\tau^2 \sum_n \int_k \epsilon_{idc} v_n^a v_n^d \left( v_n^{ib} + 2 \sum_m \epsilon_{nm}g^{ib}_{nm} \right) f_n',
\end{align*}
where the second term is determined by the conventional QMQ~\cite{QMQ1, QMQ2, QMQ3} $\epsilon_{idc}v_n^av_n^d g^{ib}_{nm}$.
Notably, for a two-dimensional Rashba electron gas with $H_0=k^2/2m+\lambda_R(k_y\hat{\sigma}^x-k_x\hat{\sigma}^y)$,
where $k^2=k_x^2+k_y^2$, we have $v_n^{ab}=0$ when $a \neq b$
and hence the OHE is entirely dominated by the conventional QMQ,
in contrast to the conventional wisdom.
The quantum geometric origin of OHE is included in Fig.~\ref{FIG1}(b).

By taking a close look at Fig.~\ref{FIG1}(b), we find that
the MNHE displays a quantum geometric duality~\cite{xiangDHE} with the PHE,
for both the spin Zeeman coupling and the orbital minimal coupling,
similar to that between the INHE and ENHE in Fig.~\ref{FIG1}(a).
Notably, all quantum geometric quantities appearing in Fig.~\ref{FIG1}(b) are gauge-invariant
and hence the corresponding response tensors can be used to evaluate the bilinear current
in realistic quantum materials when combined with first-principles calculations~\cite{Yanfirst, Fermisurf}.

To close this section, we mention that the MNHE has been studied within semiclassical theory,
where the conventional QMQ and the ZQMD normalized by energy difference
[i.e. $2\sum_{m}\mathcal{Q}^{ab}_{nm} v_n^c/\epsilon_{nm}$]
are identified as the anomalous orbital and spin polarizability dipole~\cite{MNHE1}, respectively.
In addition, the orbital-induced PHE has been explored in Weyl semimetals~\cite{PHE, MaDaPRB, OrtixPHE, YaoPHE},
but its quantum geometric duality with the orbital-induced MNHE
has not been clarified, to the best of our knowledge.
Finally, we remark that the spin-induced PHE in Fig.~\ref{FIG1}(b),
particularly revealed by the Zeeman quantum geometry,
has remained unexplored and will be discussed after
clarifying the symmetry constraints of the quantum geometric quantities appearing in Fig.~\ref{FIG1}(b).

\begin{center}
\begin{table}
\caption{\label{tab1}
{
The constraints of the ZBCD, the ZQMD, the conventional BCQ, the QMQ,
and the SBC under $\mathcal{P}$, $\mathcal{T}$, and $\mathcal{P}\mathcal{T}$ symmetries.
Here \cmark (\xmark) stands for the even (odd) parity under the assigned symmetry operation.
}
}
\begin{tabular}{ c | c |  c |  c |  c |  c}
\hline
\hline
                                         & \  ZBCD \  & \ ZQMD \ & \  BCQ \   & \ QMQ \      & \ SBC \ \ 
\\
\hline
\ \ \ $\mathcal{P}$ \ \ \                & \ \cmark \       &  \ \cmark \    & \ \cmark \  & \ \cmark \  & \ \cmark \ \
\\
\hline
\ \ \ $\mathcal{T}$ \ \ \                & \ \xmark \       &  \ \cmark \    & \ \xmark \  & \ \cmark \  & \ \cmark \ \
\\
\hline
\ \ \ $\mathcal{P}\mathcal{T}$ \ \ \     & \ \xmark \       &  \ \cmark \    & \ \xmark \  & \ \cmark \  & \ \cmark \ \
\\
\hline
\hline
\end{tabular}
\end{table}
\end{center}

\bigskip
\noindent{\textit{\textcolor{blue}{More symmetry constraints from quantum geometry}}}---Beyond
the $\mathcal{T}$-symmetry already incorporated via Onsager reciprocity,
we note that the quantum geometric quantities appearing in Fig.~\ref{FIG1}(b)
encode additional symmetry constraints for Eq.~(\ref{EBresponse})
and thereby regulate the material platforms that support the MNHE and PHE.
For example, under inversion symmetry $\mathcal{P}$,
using~\cite{xiangZeeman} $\mathcal{P}\mathcal{Z}^{ab}_{nm}=-\mathcal{Z}^{ab}_{nm}$,
$\mathcal{P}\mathcal{Q}^{ab}_{nm}=-\mathcal{Q}^{ab}_{nm}$,
$\mathcal{P}g^{ab}_{nm}=g^{ab}_{nm}$, $\mathcal{P}\Omega^{ab}_{nm}=\Omega^{ab}_{nm}$,
and $\mathcal{P}v_n^c=-v_n^c$, we find that the ZBCD $\mathcal{Z}_{nm}^{ab}v_n^c$,
the ZQMD $\mathcal{Q}_{nm}^{ab}v_n^c$,
the SBC $\Omega_{nm}^{ab}\epsilon_{cij}\Omega_{nm}^{ij}$, 
the BCQ $v_n^av_n^b\epsilon_{cij}\Omega_{nm}^{ij}$,
and the QMQ $v_n^av_n^b\epsilon_{cij}g_{nm}^{ij}$ are $\mathcal{P}$-even,
as listed in Table~\ref{tab1},
together with the constraints from $\mathcal{T}$-symmetry and the combined $\mathcal{P}\mathcal{T}$-symmetry.
Consequently, the MNHE and PHE in Eq.~(\ref{EBresponse})
can appear in both centrosymmetric and noncentrosymmetric materials,
in stark contrast to the $\mathcal{P}$-odd~\cite{NagaosaNC} nonlinear conductivity $\sigma_{abc}^{(i)}$ in Eq.~(\ref{EEresponse}),
which is expected only in noncentrosymmetric materials.

Beyond $\mathcal{P}$, $\mathcal{T}$, and $\mathcal{P}\mathcal{T}$ symmetries,
we can adopt Jahn's notation~\cite{BCQ1}
$ae[V^2]V$ and $e\{V^2\}V$ for $\sigma_{ab,c}^{(1)}$ and $\sigma_{ab,c}^{(0)}$, respectively,
and use the Bilbao Crystallographic Server~\cite{Bilbao} to
enumerate all the magnetic point groups that allow the MNHE and PHE.
Essentially, Jahn notation implements the Neumann's principle~\cite{Neumann}
for the rank-3 pseudotensor $\sigma_{ab,c}^{(i)}$ in Eq.~(\ref{EBresponse}):
$\sigma_{ab,c}^{(i)} = \eta_T |\mathcal{R}| \mathcal{R}_{aa'} \mathcal{R}_{bb'} \mathcal{R}_{cc'} \sigma_{a'b',c'}^{(i)}$,
where $\mathcal{R}_{aa'}$ is the matrix element of the point group operation $\mathcal{R}$ with
$|\mathcal{R}|$ its determinant, and $\eta_T = +1$ $(-1)$ for $\mathcal{R}$ ($\mathcal{R}\mathcal{T}$) operations.
In addition, $\{\cdots\}$ ($[\cdots]$) enforces the antisymmetric (symmetric)
permutation symmetry of $\sigma_{ab,c}^{(0)}$ ($\sigma_{ab,c}^{(1)}$) under $a \leftrightarrow b$,
as constrained by Onsager reciprocity.

In the following section, we focus on the previously unexplored spin-induced PHE.
Under the same symmetry constraint, the spin-induced PHE can generally coexist with the orbital-induced PHE.
However, for a spin-orbit-coupled two-dimensional system in an in-plane magnetic field,
the orbital-induced PHE from the orbital minimal coupling $\vect{B}\cdot(\hat{\vect{r}}\times\hat{\vect{v}})$ is suppressed,
whereas the spin-induced PHE from the spin Zeeman coupling $\vect{B}\cdot\hat{\vect{\sigma}}$ can survive.
Remarkably, most two-dimensional magnetic point groups can support this spin-induced PHE~\cite{sup}:
$1$, $2$, $2'$, $m$, $m'$, $m'm2'$, $3$, $3m$, $3m'$, $6'$, and $6'mm'$.
Guided by the symmetry analysis, we next investigate the spin-induced PHE in the
surface Dirac cone of three-dimensional TIs.

\begin{figure}[t!]
\includegraphics[width=0.95\columnwidth]{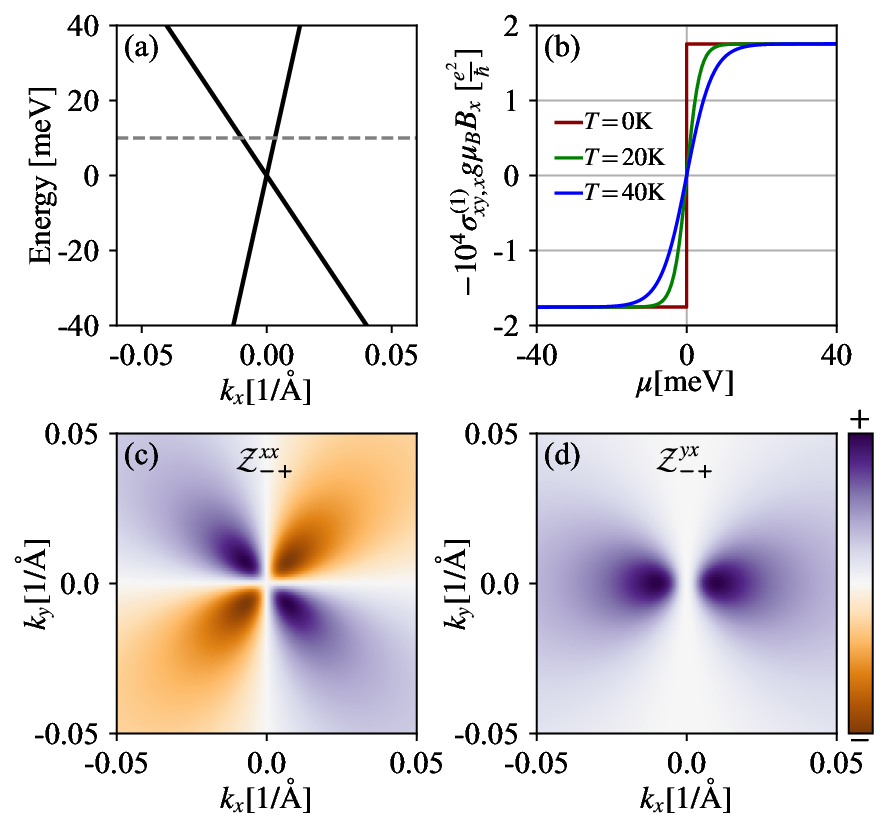}
\caption{
(a) Band dispersions of Eq.~(\ref{TIsurf}).
Here the horizontal dashed line denotes the chemical potential $\mu$.
(b) The step-like spin-induced PHE conductivity at different temperatures.
The $\vect{k}$-resolved Zeeman Berry curvature (c) $\mathcal{Z}_{-+}^{xx}$ and (d) $\mathcal{Z}_{-+}^{yx}$.
Parameters: $\beta=1/8$~\cite{SBZhangPRL}, $B_x=1 \mathrm{T}$, $\tau=0.5 \mathrm{ps}$~\cite{BCQ1} and $g=10$~\cite{gfactor}.
}
\label{FIG2}
\end{figure}

\bigskip
\noindent{\textit{\textcolor{blue}{Spin-induced PHE}}} ---
Under an in-plane magnetic field, the surface Dirac cone of TI is tilted
and its low-energy effective Hamiltonian is given by~\cite{SBZhangPRL}
\begin{align}
H=t_x k_x + v_F(k_x\hat{\sigma}^y-k_y\hat{\sigma}^x),
\label{TIsurf}
\end{align}
where $t_x$ is the tilt parameter,
$\vect{k}=(k_x, k_y)$ is the crystal momentum,
$v_F$ is the Fermi velocity,
and $\hat{\sigma}^a$ are the Pauli matrices for spin.

The band dispersions of Eq.~\eqref{TIsurf}
are $\epsilon_{\pm}=t_x k_x \pm v_Fk$, as shown in Fig.~\ref{FIG2}(a),
where $\pm$ denotes the conduction (valence) band and $k^2=k_x^2+k_y^2$.
Note that the tilt term of Eq.~(\ref{TIsurf}) breaks $\mathcal{T}$-symmetry,
but preserves the mirror symmetry $\mathcal{M}_y$ due to~\cite{FuBCD}
$\mathcal{M}_y k_x \rightarrow k_x$, $\mathcal{M}_y k_y \rightarrow -k_y$,
$\mathcal{M}_y \hat{\sigma}_x \rightarrow -\hat{\sigma}_x$, and
$\mathcal{M}_y \hat{\sigma}_y \rightarrow \hat{\sigma}_y$.
As a result, the magnetic point group of Eq.~(\ref{TIsurf})
is $m$ and the allowed spin-induced PHE conductivities are
$\sigma_{xy;x}^{(1)}=\sigma_{yx;x}^{(1)}$,
see Table~\textcolor{cyan}{I} of the Supplementary Material~\cite{sup},
which are contributed by the Zeeman Berry curvatures
$\mathcal{Z}_{\pm\mp}^{xx}=\pm k_x k_y/k^3$ and $\mathcal{Z}_{\pm\mp}^{yx}=\mp k_x^2/k^3$,
as illustrated in Fig.~\ref{FIG2}(c) and \ref{FIG2}(d), respectively.
Using polar coordinates $(k_x, k_y)=k(\cos\theta, \sin\theta)$ and evaluating Eq.~(\ref{spin1}) at zero temperature,
we obtain~\cite{sup}
\begin{align}
\sigma_{xy,x}^{(1)} 
=
\sigma_{yx,x}^{(1)} 
=
\text{sgn}(\mu)
\frac{\tau e^2\left(\beta^2+2\sqrt{1-\beta^2}-2\right)}{4\pi\hbar^2\beta^3},
\end{align}
where $\beta \equiv t_x/v_F \in (0, 1)$ is assumed, and $e$ and $\hbar$ are restored by dimensional analysis.
Fig.~\ref{FIG2}(b) shows the dependence of $\sigma_{xy,x}^{(1)}$ on the chemical potential $\mu$.
Unlike previous Fermi-surface quantum-geometric responses~\cite{BPT1, BPT2},
we find that the spin-induced PHE conductivity
features a step-like dependence on the chemical potential at zero temperature.
This step-like behavior can survive even when the temperature is slightly increased, as shown in Fig.~\ref{FIG2}(b).

We wish to mention that the in-plane Hall effect proposed in Ref.~[\onlinecite{inplaneHE}]
arises from the conventional Berry curvature and is therefore distinct from our results.
Moreover, existing experiments~\cite{PHETI} on the PHE, particularly on the surface of TIs,
may already include the spin-induced PHE proposed in this work.
Finally, using~\cite{SBZhangPRL} $\beta=1/8$, $\tau=0.5 \mathrm{ps}$~\cite{BCQ1}, $g=10$~\cite{gfactor}, $B_x=1 \mathrm{T}$,
and $E_y=10^{4} \mathrm{V/m}$,
we estimate a spin-induced PHE voltage of $\sim 43 \mathrm{\mu V}$
for a Hall bar with the resistance~\cite{BCQ1} $\sim 10^{3} \mathrm{\Omega}$
and the lateral size $\sim 100 \mathrm{\mu m}$.
Therefore, the spin-induced PHE from the Zeeman Berry curvature dipole
can be easily detected via transport measurements performed on TI surfaces.
In addition to TI surfaces, two-dimensional magnetic materials,
such as Fe$_5$GeTe$_2$~\cite{Fe5GeTe2} and TcGeSe$_3$~\cite{TcGeSe3},
see Table \textcolor{cyan}{I} of the Supplementary Material,
can also be applied to detect the spin-induced PHE.

\bigskip
\noindent{\textit{\textcolor{blue}{Discussions}}}---Equipped with the quantum geometric map
[Fig.~\ref{FIG1}(b)] for Eq.~(\ref{EBresponse}),
we notice that the orbital-induced MNHE (PHE) offers an alternative probe of the QMQ and BCQ,
which have recently attracted considerable interest but have so far
been explored mainly in the context of third-order nonlinear Hall effects~\cite{QMQ1, QMQ2, QMQ3, BCQ1}.
Besides, we mention that the spin-induced PHE in two-dimensional systems
can be used to probe the novel Zeeman quantum geometry~\cite{xiangZeeman},
as illustrated in the surface Dirac cone of TIs.
Furthermore, the orbital-induced PHE from the $\mathcal{P}\mathcal{T}$-odd BCQ (Table~\ref{tab1})
may serve as a sensitive probe of emerging altermagnets~\cite{alter1, alter2},
which break $\mathcal{P}\mathcal{T}$-symmetry and exhibit characteristic band splitting
with weak (or even negligible) spin-orbit coupling.
For example, in a planar altermagnet~\cite{QMQ2} with magnetic point group $4'/m$
the intrinsic anomalous Hall effect usually used to detect ferromagnetism is forbidden,
whereas the orbital-induced PHE (such as $\sigma_{xy;x}^{(1)}$) is symmetry-allowed.
Recently, Ohm's law has been revisited by considering the nonlinear charge current in Eq.~(\ref{EEresponse}),
particularly in noncentrosymmetric materials~\cite{Ohm, Ohm1}.
Under electromagnetic fields, however, this law has so far accounted only for the OHE~\cite{Ohm}
and thus should also be revisited by properly considering the spin-induced (orbital-induced) MNHE and PHE~\cite{GaoY2014PRL,MNHE2}.

In addition to the charge degree of freedom,
we expect that similar quantum geometric maps can be developed for other degrees of freedom in solids,
such as spin~\cite{xiangSHE}, orbital~\cite{orbital0, orbital1, orbital2, orbital3, orbital4, orbital5},
layer~\cite{layer1}, and valley~\cite{valley0, valley1, valley2, valley3, valley4}.
We close by remarking that our quantum geometric map is bulit on the relaxation time approximation,
where the relaxation time $\tau$ is introduced phenomenologically.
Beyond this approximation, however,
a microscopic treatment~\cite{B1, B2} where the relaxation arises from the explicit coupling to a bath,
such as in Floquet systems~\cite{F1, F2, F3}, is necessary,
and whether there exists a quantum geometric map of electromagnetic responses is an interesting question.

\bigskip
\section*{Acknowledgements}
We thank the National Natural Science Foundation of China (Grants No. 12404059, No.12574054, and No. 12034014).


\begin{thebibliography}{00}
\bibitem{Xiao2010}
D. Xiao, M.-C. Chang, and Q. Niu,
Berry phase effects on electronic properties,
\href{https://doi.org/10.1103/RevModPhys.82.1959}{Rev. Mod. Phys. \textbf{82}, 1959 (2010).}
\bibitem{QuantumGeometry}
P\"aivi T\"orm\"a,
Essay: Where can quantum geometry lead us?,
\href{https://doi.org/10.1103/PhysRevLett.131.240001}
{Phys. Rev. Lett. \textbf{131}, 240001 (2023).}
\bibitem{YanPRR}
T. Holder, D. Kaplan, and B. Yan,
Consequences of time-reversal-symmetry breaking in the light-matter interaction:
Berry curvature, quantum metric, and diabatic motion,
\href{https://doi.org/10.1103/PhysRevResearch.2.033100}
{Phys. Rev. Research \textbf{2}, 033100 (2020).}
\bibitem{QuantumGeometry1}
J. Ahn, G.-Y. Guo, N. Nagaosa,
Low-frequency divergence and quantum geometry of the bulk photovoltaic effect in topological semimetals,
\href{https://doi.org/10.1103/PhysRevX.10.041041}
{Phys. Rev. X \textbf{10}, 041041 (2020).}
\bibitem{QuantumGeometry2}
J. Ahn, G.-Y. Guo, N. Nagaosa, and A. Vishwanath,
Riemannian geometry of resonant optical responses,
\href{https://doi.org/10.1038/s41567-021-01465-z}
{Nat. Phys. \textbf{18}, 290 (2022).}
\bibitem{MaQlight}
Q. Ma, A. G. Grushin and K. S. Burch,
Topology and geometry under the nonlinear electromagnetic spotlight,
\href{https://doi.org/10.1038/s41563-021-00992-7}
{Nat. Mater. \textbf{20}, 1601 (2021).}
\bibitem{JTsong}
Q. Ma, R. K. Kumar, S.-Y. Xu, F. H. L. Koppens, and J. C. W. Song,
Photocurrent as a multiphysics diagnostic of quantum materials,
\href{https://doi.org/10.1038/s42254-022-00551-2}
{Nat. Rev. Phys. \textbf{5}, 170 (2023).}
\bibitem{AMIT}
P. C. Adak, S. Sinha, A. Agarwal, and M. M. Deshmukh,
Tunable moir\'e materials for probing Berry physics and topology,
\href{https://doi.org/10.1038/s41578-024-00671-4}
{Nat. Rev. Mater. \textbf{9}, 481 (2024).}
\bibitem{XiaoCQF}
J. Li, D. Zhai, C. Xiao, and W. Yao,
Dynamical chiral Nernst effect in twisted Van der Waals few layers,
\href{https://doi.org/10.1007/s44214-024-00059-z}{Quantum Front. \textbf{3}, 11 (2024)}.
\bibitem{Rashbagas}
G. Sala, M. T. Mercaldo, K. Domi, S. Gariglio, M. Cuoco, C. Ortix, and A. D. Caviglia,
The quantum metric of electrons with spin-momentum locking,
\href{https://doi.org/10.1126/science.adq3255}{Science \textbf{389}, 822 (2025)}.
\bibitem{entropy}
T. B. Smith, L. Pullasseri, and A. Srivastava,
Momentum-space gravity from the quantum geometry and entropy of Bloch electrons,
\href{https://doi.org/10.1103/PhysRevResearch.4.013217}{Phys. Rev. Research \textbf{4}, 013217 (2022)}.
\bibitem{PRA}
B. Het\'enyi and P\'eter L\'evay,
Fluctuations, uncertainty relations, and the geometry of quantum state manifolds,
\href{https://doi.org/10.1103/PhysRevA.108.032218}{Phys. Rev. A \textbf{108}, 032218 (2023).}
\bibitem{Berniverg}
J. B. Yu, B. A. Bernevig, R. Queiroz, E. Rossi, P. T\"{o}rm\"{a}, and B.-J. Yang,
\href{https://doi.org/10.1038/s41535-025-00801-3}{Quantum Mater. \textbf{10}, 101 (2025)}.
\bibitem{YanReview}
Y. Jiang, T. Holder, and B. H. Yan,
Revealing quantum geometry in nonlinear quantum materials,
\href{https://doi.org/10.1088/1361-6633/ade454}{Rep. Prog. Phys. \textbf{88}, 076502 (2025)}.
\bibitem{HZLu2024}
T. Liu, X.-B. Qiang, H.-Z. Lu, X. C. Xie,
Quantum geometry in condensed matter,
\href{https://doi.org/10.1093/nsr/nwae334}{Nat. Sci. Rev. \textbf{12}, nwae334 (2024).}
\bibitem{Fontana}
P. Fontana, V. Velasco, C. Niu, P. D. Ye, P. V. Lopes, K. E. M. de Souza, M. V. O. Moutinho, C. Lewenkopf, and M. B. Silva Neto,
Quantum Geometry and the Electric Magnetochiral Anisotropy in Noncentrosymmetric Polar Media,
\href{https://doi.org/10.1103/7nxc-j62y}{Phys. Rev. Lett. \textbf{135}, 106602 (2025)}.
\bibitem{SuYangXuRMP}
A. Gao, N. Nagaosa, N. Ni, S.-Y. Xu,
Quantum Geometry Phenomena in Condensed Matter Systems,
Preprint at \href{https://doi.org/10.48550/arXiv.2508.00469}{https://doi.org/10.48550/arXiv.2508.00469}.
\bibitem{DaoDao}
H. Wang and K. Chang,
Geodesic nature and quantization of shift vector,
Preprint at
\href{https://doi.org/10.48550/arXiv.2405.13355}{https://doi.org/10.48550/arXiv.2405.13355}.
\bibitem{QueirozReview}
N. Verma, P. J. W. Moll, T. Holder, and R. Queiroz,
Quantum Geometry: Revisiting electronic scales in quantum matter,
Preprint at \href{https://doi.org/10.48550/arXiv.2504.07173}{https://doi.org/10.48550/arXiv.2504.07173}.
\bibitem{xiangSHE}
L. Xiang, H. Jin, and J. Wang,
Spin transport revealed by the spin quantum geometry,
\href{https://doi.org/10.1103/14mp-263q}{Phys. Rev. Lett. \textbf{135}, 146303 (2025)}.
\bibitem{GaoYFOP}
Y. Gao,
Semiclassical dynamics and nonlinear charge current,
\href{https://doi.org/10.1007/s11467-019-0887-2}{Front. Phys. \textbf{14}, 33404 (2019)}.
\bibitem{tau}
The dependence of $\sigma^{(i)}_{abc}$ and $\sigma^{(i)}_{ab,c}$ on the relaxation time $\tau$
should not be taken as a general criterion for the time-reversal character of the response,
which is instead determined by the underlying microscopic quantum geometric origin.
\bibitem{GaoY2014PRL}
Y. Gao, S. Y. A. Yang, and Q. Niu,
Field Induced Positional Shift of Bloch Electrons and Its Dynamical Implications,
\href{https://doi.org/10.1103/PhysRevLett.112.166601}
{Phys. Rev. Lett. \textbf{112}, 166601 (2014)}.
\bibitem{BPT1}
C. Wang, Y. Gao, and D. Xiao,
Intrinsic nonlinear Hall effect in antiferromagnetic tetragonal CuMnAs,
\href{https://doi.org/10.1103/PhysRevLett.127.277201}
{Phys. Rev. Lett. \textbf{127}, 277201 (2021).}
\bibitem{BPT2}
H. Liu, J. Zhao, Y.-X. Huang, W. Wu, X.-L. Sheng, C. Xiao, and S. Y. A. Yang,
Intrinsic second-order anomalous Hall effect and its application in compensated antiferromagnets,
\href{https://doi.org/10.1103/PhysRevLett.127.277202}
{Phys. Rev. Lett. \textbf{127}, 277202 (2021)}.
\bibitem{YanPRL2024}
D. Kaplan, T. Holder, and B.-H. Yan,
Unification of Nonlinear Anomalous Hall Effect and Nonreciprocal Magnetoresistance in Metals by the Quantum Geometry,
\href{https://doi.org/10.1103/PhysRevLett.132.026301}
{Phys. Rev. Lett. \textbf{132}, 026301 (2024)}.
\bibitem{CulcerPRBL}
K. Das, S. Lahiri, R. B. Atencia, D. Culcer, and A. Agarwal,
Intrinsic nonlinear conductivities induced by the quantum metric,
\href{https://doi.org/10.1103/PhysRevB.108.L201405}
{Phys. Rev. B \textbf{108}, L201405 (2023).}
\bibitem{GangSu2022}
Y. Wang, Z. Zhang, Z.-G. Zhu, and G. Su,
Intrinsic nonlinear Ohmic current,
\href{https://doi.org/10.1103/PhysRevB.109.085419}
{Phys. Rev. B \textbf{109}, 085419 (2024).}
\bibitem{Jia2024}
J. Jia, L. Xiang, Z. Qiao, and J. Wang, 
Equivalence of semiclassical and response theories for second-order nonlinear ac Hall effects,
\href{https://doi.org/10.1103/PhysRevB.110.245406}{Phys. Rev. B \textbf{110}, 245406 (2024).}
\bibitem{XuSY2023}
A. Gao, Y.-F. Liu, J.-X. Qiu, B. Ghosh, T. V. Trevisan, Y. Onishi, C. Hu, T. Qian, H.-J. Tien, S.-W. Chen, \textit{et al.},
Quantum metric nonlinear Hall effect in a topological antiferromagnetic heterostructure,
\href{https://doi.org/10.1126/science.adf1506}
{Science \textbf{381}, 181 (2023).}
\bibitem{Wang2023}
N. Wang, D. Kaplan, Z. Zhang, T. Holder, N. Cao, A. Wang, X.  Zhou, F. Zhou, Z. Jiang, C. Zhang \textit{et al.},
Quantum-metric-induced nonlinear transport in a topological antiferromagnet,
\href{https://doi.org/10.1038/s41586-023-06363-3}
{Nature (London) \textbf{621}, 487 (2023).}
\bibitem{Han2024}
J. Han, T. Uchimura, Y. Araki, J.-Y. Yoon, Y. Takeuchi, Y. Yamane, S. Kanai, J. Ieda, H. Ohno, and S. Fukami,
Room-temperature flexible manipulation of the quantum-metric structure in a topological chiral antiferromagnet,
\href{https://doi.org/10.1038/s41567-024-02476-2}{Nat. Phys. \textbf{20}, 1110 (2024).}
\bibitem{FuBCD}
I. Sodemann and L. Fu,
Quantum nonlinear Hall effect induced by Berry curvature dipole in time-reversal invariant materials,
\href{https://doi.org/10.1103/PhysRevLett.115.216806}
{Phys. Rev. Lett. \textbf{115}, 216806 (2015).}
\bibitem{BCDexp1}
Q. Ma, S.-Y. Xu, H. Shen, D. MacNeill, V. Fatemi, T.-R. Chang,
A. M. M. Valdivia, S. Wu, Z. Du, C.-H. Hsu, S. Fang, Q. D. Gibson,
K. Watanabe, T. Taniguchi, R. J. Cava, E. Kaxiras,
H.-Z. Lu, H. Lin, L. Fu, N. Gedik, and P. Jarillo-Herrero,
Observation of the nonlinear Hall effect under time-reversal-symmetric conditions,
\href{https://doi.org/10.1038/s41586-018-0807-6}
{Nature \textbf{565}, 337 (2019).}
\bibitem{BCDexp2}
K. Kang, T. Li, E. Sohn, J. Shan, and K. F. Mak,
Nonlinear anomalous Hall effect in few-layer WTe$_2$,
\href{https://doi.org/10.1038/s41563-019-0294-7}
{Nat. Mater. \textbf{18}, 324 (2019).}
\bibitem{BCDexp3}
D. Kumar, C.-H. Hsu, R. Sharma, T.-R. Chang, P. Yu, J. Wang, G. Eda, G. Liang, and H. Yang,
Room-temperature nonlinear Hall effect and wireless radiofrequency rectification in Weyl semimetal TaIrTe$_4$,
\href{https://doi.org/10.1038/s41565-020-00839-3}
{Nat. Nanotechnol. \textbf{16}, 421 (2021).}
\bibitem{QMDfootnote}
The nonlinear Drude conductivity~\cite{FuBCD} is given by
$\sigma_{abc}^{(2)}=\tau^2 \sum_n \int_k v_n^a \partial_{bc}^2f_n$,
using $\partial_b v_n^a = v_{n}^{ab} + 2 \sum_{m}\epsilon_{nm}g^{ab}_{nm}$,
where $v_n^a=v_{nn}^a$ and $v_{n}^{ab} \equiv \langle u_n|\partial^2_{ab}H|u_n\rangle$
and $g^{ab}_{nm} \equiv \left(r^a_{nm}r^b_{mn}+r^b_{nm}r^a_{mn}\right)/2$ (quantum metric),
by integration by parts, we find that $\sigma_{abc}^{(2)}$ includes an interband contribution
$-2 \tau^2 \sum_m \int_k \epsilon_{nm}g^{ab}_{nm}v_n^cf_n'$,
which is governed by the quantum metric dipole $g^{ab}_{nm}v_n^c$.
\bibitem{NagaosaDrude}
K. Shinada and N. Nagaosa,
Quantum geometric bounds for observables: Linear responses, Drude weight, and orbital magnetization,
\href{https://doi.org/10.1103/qxbl-qd4f}{Phys. Rev. B \textbf{112}, 155158 (2025)}.
\bibitem{Psym}
For example, in addition to the $\mathcal{T}$-symmetry,
we note that both the Berry curvature dipole and quantum metric dipole are $\mathcal{P}$-odd
with $\mathcal{P}$ the inversion symmetry, which dictates that the nonlinear currents in Eq.~\eqref{EEresponse}
can only appear in noncentrosymmetric systems.
\bibitem{PHE}
S. Nandy, G. Sharma, A. Taraphder, and S. Tewari,
Chiral Anomaly as the Origin of the Planar Hall Effect in Weyl Semimetals,
\href{https://doi.org/10.1103/PhysRevLett.119.176804}{Phys. Rev. Lett. \textbf{119}, 176804 (2017).}
\bibitem{MaDaPRB}
D. Ma, H. Jiang, H. W. Liu, and X. C. Xie,
Planar Hall effect in tilted Weyl semimetals,
\href{https://doi.org/10.1103/PhysRevB.99.115121}{Phys. Rev. B \textbf{99}, 115121 (2019).}
\bibitem{OrtixPHE}
R. Battilomo, N. Scopigno, and C. Ortix,
Anomalous planar Hall effect in two-dimensional trigonal crystals,
\href{https://doi.org/10.1103/PhysRevResearch.3.L012006}{Phys. Rev. Research \textbf{3}, L012006 (2021).}
\bibitem{YaoPHE}
L. Li, J. Cao, C. Cui, Z.-M. Yu, and Y.-G. Yao,
Planar Hall effect in topological Weyl and nodal-line semimetals,
\href{https://doi.org/10.1103/PhysRevB.108.085120}{Phys. Rev. B \textbf{108}, 085120 (2023)}
\bibitem{MNHE1}
H. Wang, Y. Huang, H. Liu, X. Feng, J. Zhu, W. Wu, C. Xiao, and S. Y. A. Yang,
Orbital Origin of the Intrinsic Planar Hall Effect,
\href{https://doi.org/10.1103/PhysRevLett.132.056301}{Phys. Rev. Lett. \textbf{132}, 056301 (2024).}
\bibitem{MNHE2}
Z. Du, Y.-X. Huang, and X. Li,
Orbital and spin bilinear magnetotransport effect in Weyl/Dirac semimetal,
\href{https://doi.org/10.1103/npvn-33r8}{Phys. Rev. B \textbf{112}, 064201 (2025)}.
\bibitem{MNHE3}
L. Xiang and J. Wang,
Intrinsic in-plane magnetononlinear Hall effect in tilted Weyl semimetals,
\href{https://doi.org/10.1103/PhysRevB.109.075419}{Phys. Rev. B \textbf{109}, 075419 (2024).}
\bibitem{MNHE4}
L. Wang, J. Zhu, H. Chen, H. Wang, J. Liu, Y.-X. Huang,
B. Jiang, J. Zhao, H. Shi, G. Tian, H. Wang, Y.-G. Yao, D.-P Yu, Z. Wang, C. Xiao, S. Y. A. Yang, and X. S. Wu,
Orbital magneto-nonlinear anomalous Hall effect in kagome magnet Fe$_3$Sn$_2$,
\href{https://doi.org/10.1103/PhysRevLett.132.106601}{Phys. Rev. Lett. \textbf{132}, 106601 (2024).}
\bibitem{MNHE5}
Y. Wang, Z.-G. Zhu, and G. Su,
Field-induced Berry connection and anomalous planar Hall effect in tilted Weyl semimetals,
\href{https://doi.org/10.1103/PhysRevResearch.5.043156}{Phys. Rev. Research \textbf{5}, 043156 (2023).}
\bibitem{Nagaosa2010}
N. Nagaosa, J. Sinova, S. Onoda, A. H. MacDonald, and N. P. Ong,
Anomalous Hall effect,
\href{https://doi.org/10.1103/RevModPhys.82.1539}
{Rev. Mod. Phys. \textbf{82}, 1539 (2010).}
\bibitem{xiangZeeman}
L. Xiang, J. Jia, F. Xu, Z. Qiao, and J. Wang,
Intrinsic Gyrotropic Magnetic Current from Zeeman Quantum Geometry,
\href{https://doi.org/10.1103/PhysRevLett.134.116301}{Phys. Rev. Lett. \textbf{134}, 116301 (2025).}
\bibitem{Zeeman1}
J. Cao, F. Qi, Y. Xiang, and G. Jin,
Magnetoelectric response arising from Zeeman quantum geometry,
\href{https://doi.org/10.1103/54qq-2qgh}{Phys. Rev. B \textbf{112}, 235145 (2025)}.
\bibitem{Zeeman2}
M. Ezawa,
Quantum geometry and X-wave magnets with X=p,d,f,g,i,
Preprint at \href{https://doi.org/10.48550/arXiv.2512.05477}{https://doi.org/10.48550/arXiv.2512.05477}.
\bibitem{Zeeman3}
J. Tan, O. Matsyshyn, G. Vignale, and J. C. W. Song,
Nonequilibrium Exchange Nonlinear Hall Effect,
Preprint at \href{https://doi.org/10.48550/arXiv.2512.06074}{https://doi.org/10.48550/arXiv.2512.06074}.
\bibitem{Zeeman4}
N. Chakraborti, S. K. Ghosh, and S. Nandy,
Zeeman Quantum Geometry as a Probe of Unconventional Magnetism,
Preprint at \href{https://doi.org/10.48550/arXiv.2508.14745}{https://doi.org/10.48550/arXiv.2508.14745}.
\bibitem{QMQ1}
H. Li, C. Zhang, C. Zhou, C. Ma, X. Lei, Z. Jin, H. He, B. Li, K. T. Law, and J. Wang,
Quantum geometry quadrupole-induced third-order nonlinear transport in antiferromagnetic topological insulator MnBi$_2$Te$_4$,
\href{https://doi.org/10.1038/s41467-024-52206-8}{Nat. Commun. \textbf{15}, 7779 (2024).}
\bibitem{QMQ2}
Y. Fang, J. Cano, and S. A. A. Ghorashi,
Quantum Geometry Induced Nonlinear Transport in Altermagnets,
\href{https://doi.org/10.1103/PhysRevLett.133.106701}{Phys. Rev. Lett. \textbf{133}, 106701 (2024)}.
\bibitem{QMQ3}
X. Liu, A. Wang, D. Li, T. Zhao, X. Liao, and Z. M. Liao,
Giant Third-Order Nonlinearity Induced by the Quantum Metric Quadrupole in Few-Layer WTe$_2$,
\href{https://doi.org/10.1103/PhysRevLett.134.026305}{Phys. Rev. Lett. \textbf{134}, 026305 (2025).}
\bibitem{Duan2024}
J. J. Yao, Y. Z. Liu, and W. H. Duan,
Geometrical nonlinear Hall effect induced by Lorentz force,
\href{https://doi.org/10.1103/PhysRevB.110.115123}{Phys. Rev. B \textbf{110}, 115123 (2024).}
\bibitem{BCQ1}
C.-P. Zhang, X.-J. Gao, Y.-M. Xie, H. C. Po, and K. T. Law,
Higher-order nonlinear anomalous Hall effects induced by Berry curvature multipoles,
\href{https://doi.org/10.1103/PhysRevB.107.115142}{Phys. Rev. B \textbf{107}, 115142 (2023).}
\bibitem{BCQ2}
D. E. Parker, T. Morimoto, J. Orenstein, and J. E. Moore,
Diagrammatic approach to nonlinear optical response with application to Weyl semimetals,
\href{https://doi.org/10.1103/PhysRevB.99.045121}{Phys. Rev. B \textbf{99}, 045121 (2019)}.
\bibitem{BCQ3}
S. Sankar, R. Liu, C.-P. Zhang, Q.-F. Li, C. Chen, X.-J. Gao, J. Zheng, Y.-H. Lin, K. Qian, R.-P. Yu, X. Zhang, Z. Y. Meng,
K. T. Law, Q. Shao, and B. J\"{a}ck,
Experimental Evidence for a Berry Curvature Quadrupole in an Antiferromagnet,
\href{https://doi.org/10.1103/PhysRevX.14.021046}{Phys. Rev. X \textbf{14}, 021046 (2024)}.
\bibitem{Onsager1}
L. Onsager,
Reciprocal Relations in Irreversible Processes. I.,
\href{https://doi.org/10.1103/PhysRev.37.405}{Phys. Rev. \textbf{37}, 405 (1931)}.
\bibitem{Onsager2}
L. Onsager,
Reciprocal Relations in Irreversible Processes. II.,
\href{https://doi.org/10.1103/PhysRev.38.2265}{Phys. Rev. \textbf{38}, 2265 (1931)}.
\bibitem{Onsager3}
H. B. G. Casimir,
On Onsager's Principle of Microscopic Reversibility,
\href{https://doi.org/10.1103/RevModPhys.17.343}{Rev. Mod. Phys. \textbf{17}, 343 (1945)}.
\bibitem{anticomponent}
For the antisymmetric component,
we have $\sigma_{ab}^{A}(0,0)=-\sigma_{ab}^{A}(0,0)=0$ when $\vect{M}=0$ and $\vect{B}=0$.
\bibitem{Sipe0}
C. Aversa and J. E. Sipe,
Nonlinear optical susceptibilities of semiconductors: Results with a length-gauge analysis,
\href{https://doi.org/10.1103/PhysRevB.52.14636}{Phys. Rev. B \textbf{52}, 14636 (1995).}
\bibitem{Sipe1}
J. E. Sipe and A. I. Shkrebtii,
Second-order optical response in semiconductors,
\href{https://doi.org/10.1103/PhysRevB.61.5337}{Phys. Rev. B \textbf{61}, 5337 (2000).}
\bibitem{sup}
Supplementary Material,
which includes Refs.~\onlinecite{Xiao2010, GaoY2014PRL, Jia2024, xiangZeeman,MNHE1,QMQ2, Sipe0, Sipe1, Fe5GeTe2,
VSe2,TcGeSe3,MnPS3,NbSi2N4}.
\bibitem{Fe5GeTe2}
Z. Guo, S. Qian, X. Zhou, W. Wang, Z. Cheng, and X. Wang,
Sliding ferroelectric metal with ferrimagnetism,
\href{https://doi.org/10.1038/s41467-025-67240-3}{Nat. Commun. \textbf{17}, 549 (2026)}.
\bibitem{VSe2}
F. Wang, Y. Zhou, X. Shen, S. Dong, and J. Zhan,
Magnetoelectric coupling and cross control in two-dimensional ferromagnets,
\href{https://doi.org/10.1103/PhysRevApplied.20.064011}{Phys. Rev. Applied \textbf{20}, 064011 (2023)}.
\bibitem{TcGeSe3}
Y. Zhu, J. Sun, J. Pan, J. Deng, and S. Du,
Enforced Symmetry Breaking for Anomalous Valley Hall Effect in Two-Dimensional Hexagonal Lattices,
\href{https://doi.org/10.1103/PhysRevLett.134.046403}{Phys. Rev. Lett. \textbf{134}, 046403 (2025)}.
\bibitem{MnPS3}
S. N. Neal, H.-S. Kim, K. A. Smith, A. V. Haglund, D. G. Mandrus, H. A. Bechtel,
G. L. Carr, K. Haule, D. Vanderbilt, and J. L. Musfeldt,
Near-field infrared spectroscopy of monolayer MnPS$_3$,
\href{https://doi.org/10.1103/PhysRevB.100.075428}{Phys. Rev. B \textbf{100}, 075428 (2019)}.
\bibitem{NbSi2N4}
Y. Yang, Y. Zhao, F. Kong, Q. Liu, Y. Su, J. Zhao, and X. Jiang,
Interlayer Sliding Induced Triferroic Coupling in 2D Bilayer NbSi$_2$N$_4$,
\href{https://doi.org/10.1021/acs.jpclett.5c02112}{J. Phys. Chem. Lett. \textbf{16}, 8649 (2025)}.
\bibitem{BMRfoot}
Eq.~\eqref{spin1} can also be responsible for the bilinear magnetoresistance~\cite{HePanNP, BMR} when $a=b$.
\bibitem{HePanNP}
P. He, S. S.-L. Zhang, D. Zhu, Y. Liu, Y. Wang, J. Yu, G. Vignale, and H. Yang,
Bilinear magnetoelectric resistance as a probe of three-dimensional spin texture in topological surface states,
\href{https://doi.org/10.1038/s41567-017-0039-y}{Nat. Phys. \textbf{14}, 495 (2018)}.
\bibitem{BMR}
D. Kim, K. Kim, K. Lee, J. H. Oh, X. Chen, S. Yang, Y. Pu, Y. Liu, F. Hu, P. C. Van, J. Jeong, K. Lee, and H. Yang,
Spin Hall-induced bilinear magnetoelectric resistance,
\href{https://doi.org/10.1038/s41563-024-02000-0}{Nat. Mater. \textbf{23}, 1509 (2024)}
\bibitem{xiangDHE}
L. Xiang, B. Wang, Y. Wei, Z. Qiao, and J. Wang,
Linear displacement current solely driven by the quantum metric,
\href{https://journals.aps.org/prb/abstract/10.1103/PhysRevB.109.115121}
{Phys. Rev. B \textbf{109}, 115121 (2024).}
\bibitem{Yanfirst}
J. Xiao and B.-H. Yan,
First-principles calculations for topological quantum materials,
\href{https://www.nature.com/articles/s42254-021-00292-8}
{Nat. Rev. Phys. \textbf{3}, 283 (2021).}
\bibitem{Fermisurf}
Note that the spin-induced MNHE and PHE feature a Fermi-surface property~\cite{Xiao2010}
due to the presence of $f_n'$ in Eqs.~(\ref{spin0}-\ref{spin1}),
and can only appear in systems with a finite Fermi surface,
which is the case of the orbital-induced MNHE and PHE.
\bibitem{NagaosaNC}
Y. Tokura and N. Nagaosa,
Nonreciprocal responses from non-centrosymmetric quantum materials,
\href{https://doi.org/10.1038/s41467-018-05759-4}{Nat. Commun. \textbf{9}, 3740 (2018).}
\bibitem{Bilbao}
S. V. Gallego, J. Etxebarria, L. Elcoro, E. S. Tasci, and J. M. Perez-Mato,
Automatic calculation of symmetry-adapted tensors in magnetic and non-magnetic materials:
a new tool of the Bilbao Crystallographic Server,
\href{https://doi.org/10.1107/S2053273319001748}{Acta Crystallogr. Sect. A \textbf{75}, 438 (2019)}.
\bibitem{Neumann}
R. E. Newnham, Properties of materials: anisotropy, symmetry, structure (Oxford university press, 2005).
\bibitem{SBZhangPRL}
S. B. Zhang, C. A. Li, F. Pena-Benitez, P. Surowka, R. Moessner, L. W. Molenkamp, and B. Trauzettel,
Super-Resonant Transport of Topological Surface States Subjected to In-Plane Magnetic Fields,
\href{https://doi.org/10.1103/PhysRevLett.127.076601}{Phys. Rev. Lett. \textbf{127}, 076601 (2021)}.
\bibitem{inplaneHE}
V. A. Zyuzin,
In-plane Hall effect in two-dimensional helical electron systems,
\href{https://doi.org/10.1103/PhysRevB.102.241105}{Phys. Rev. B \textbf{102}, 241105(R) (2020)}.
\bibitem{PHETI}
A. A. Taskin, H. F. Legg, F. Yang, S. Sasaki, Y. Kanai, K. Matsumoto, A. Rosch, and Y. Ando,
Planar Hall effect from the surface of topological insulators,
\href{https://doi.org/10.1038/s41467-017-01474-8}{Nat. Commun. \textbf{8}, 1340 (2017)}.
\bibitem{gfactor}
W. Miao, B. Guo, S. Stemmer, and X. Dai,
Engineering the in-plane anomalous Hall effect in Cd$_3$As$_2$ thin films,
\href{https://doi.org/10.1103/PhysRevB.109.155408}{Phys. Rev. B \textbf{109}, 155408 (2024)}.
\bibitem{alter1}
L. \v{S}mejkal, Jairo Sinova, and T. Jungwirth,
Beyond Conventional Ferromagnetism and Antiferromagnetism: A Phase with Nonrelativistic Spin and Crystal Rotation Symmetry,
\href{https://doi.org/10.1103/PhysRevX.12.031042}{Phys. Rev. X \textbf{12}, 031042 (2022)}.
\bibitem{alter2}
L. \v{S}mejkal, Jairo Sinova, and T. Jungwirth,
Emerging Research Landscape of Altermagnetism,
\href{https://doi.org/10.1103/PhysRevX.12.040501}{Phys. Rev. X \textbf{12}, 040501 (2022)}.
\bibitem{Ohm}
M. Su\'arez-Rodr\'iguez, F. De Juan, I. Souza, M. Gobbi, F. Casanova, and L. E. Hueso,
Non-linear Transport in Non-centrosymmetric Systems: From Fundamentals to Applications,
\href{https://doi.org/10.1038/s41563-025-02261-3}{Nat. Mater. \textbf{24}, 1005 (2025)}.
\bibitem{Ohm1}
L. Min, Y. Zhang, Z. Xie, S. Venkata G. Ayyagari, L. Miao, Y. Onishi, S. H. Lee, Y. Wang, N. Alem, L. Fu, and Z. Mao,
Colossal room-temperature non-reciprocal Hall effect,
\href{https://doi.org/10.1038/s41563-024-02015-7}{Nat. Mater. \textbf{23}, 1671 (2024)}.
\bibitem{orbital0}
B. A. Bernevig, T. L. Hughes, and S.-C. Zhang,
Orbitronics: The Intrinsic Orbital Current in $p$-Doped Silicon,
\href{https://doi.org/10.1103/PhysRevLett.95.066601}{Phys. Rev. Lett. \textbf{95}, 066601 (2005)}.
\bibitem{orbital1}
Y.-G. Choi, D. Jo, K.-H. Ko, D. Go, K.-H. Kim, H. G. Park, C. Kim, B.-C. Min, G.-M. Choi, and H.-W. Lee,
Observation of the orbital Hall effect in a light metal Ti,
\href{https://doi.org/10.1038/s41586-023-06101-9}{Nature \textbf{619}, 52 (2023)}.
\bibitem{orbital2}
I. Lyalin, S. Alikhah, M. Berritta, P. M. Oppeneer, and R. K. Kawakami,
Magneto-Optical Detection of the Orbital Hall Effect in Chromium,
\href{https://doi.org/10.1103/PhysRevLett.131.156702}{Phys. Rev. Lett. \textbf{131}, 156702 (2023)}.
\bibitem{orbital3}
G. Sala, H. Wang, W. Legrand, and P. Gambardella,
Orbital Hanle Magnetoresistance in a $3d$ Transition Metal,
\href{https://doi.org/10.1103/PhysRevLett.131.156703}{Phys. Rev. Lett. \textbf{131}, 239901 (2023)}.
\bibitem{orbital4}
D. Das, Orbitronics in action,
\href{https://doi.org/10.1038/s41567-023-02183-4}{Nat. Phys. \textbf{19}, 1085 (2023)}.
\bibitem{orbital5}
R. B. Atencia, A. Agarwal, and D. Culcer,
Orbital angular momentum of Bloch electrons: equilibrium formulation, magneto-electric phenomena, and the orbital Hall effect,
\href{https://doi.org/10.1080/23746149.2024.2371972}{Advances in Physics: X \textbf{9}, 2371972 (2024)}.
\bibitem{layer1}
A. Gao, Y.-F. Liu, C. Hu, J.-X. Qiu, C. Tzschaschel, B. Ghosh, S.-C. Ho, D. B\'erub\'e, R. Chen, H. Sun, Z. Zhang,
X.-Y. Zhang, Y.-X. Wang, N. Wang, Z. Huang, C. Felser, A. Agarwal, T. Ding, H.-J. Tien, A. Akey, J. Gardener, B. Singh,
K. Watanabe, T. Taniguchi, K. S. Burch, D. C. Bell, B. B. Zhou, W. Gao, H.-Z. Lu, A. Bansil, H. Lin, T.-R. Chang, L. Fu,
Q. Ma, N. Ni, and S.-Y. Xu,
Layer Hall effect in a 2D topological axion antiferromagnet,
\href{https://doi.org/10.1038/s41586-021-03679-w}{Nature \textbf{595}, 521 (2021)}.
\bibitem{valley0}
D. Xiao, W. Yao, and Q. Niu,
Valley-Contrasting Physics in Graphene: Magnetic Moment and Topological Transport,
\href{https://doi.org/10.1103/PhysRevLett.99.236809}{Phys. Rev. Lett. \textbf{99}, 236809 (2007)}.
\bibitem{valley1}
K. F. Mak, K. L., McGill, J. Park, and P. L. McEuen,
The valley Hall effect in MoS2 transistors,
\href{https://doi.org/10.1126/science.1250140}{Science \textbf{344}, 1489 (2014)}.
\bibitem{valley2}
M. Sui, G. Chen, L. Ma, W.-Y. Shan, D. Tian, K. Watanabe, T. Taniguchi, X. Jin, W. Yao, D. Xiao, and Y. Zhang,
Gate-tunable topological valley transport in bilayer graphene,
\href{https://doi.org/10.1038/nphys3485}{Nat. Phys. \textbf{11}, 1027 (2015)}.
\bibitem{valley3}
Y. Shimazaki, M. Yamamoto, I. V. Borzenets, K. Watanabe, T. Taniguchi, and S. Tarucha,
Generation and detection of pure valley current by electrically induced Berry curvature in bilayer graphene,
\href{https://doi.org/10.1038/nphys3551}{Nat. Phys. \textbf{11}, 1032 (2015)}.
\bibitem{valley4}
K. Das, K. Ghorai, D. Culcer, and A. Agarwal,
Nonlinear Valley Hall Effect,
\href{https://doi.org/10.1103/PhysRevLett.132.096302}{Phys. Rev. Lett. \textbf{132}, 096302 (2024)}.
\bibitem{B1}
L. Shi, O. Matsyshyn, J. C. W. Song, and I. S. Villadiego,
Floquet Fermi Liquid,
\href{https://doi.org/10.1103/PhysRevLett.132.146402}{Phys. Rev. Lett. \textbf{132}, 146402 (2024)}.
\bibitem{B2}
L. Shi, O. Matsyshyn, J. C. W. Song, and I. S. Villadiego,
Ultracritical Floquet Non-Fermi Liquid,
\href{https://doi.org/10.1103/PhysRevLett.134.196401}{Phys. Rev. Lett. \textbf{134}, 196401 (2025)}.
\bibitem{F1}
I. Esin, M. S. Rudner, G. Refael, and N. H. Lindner,
Quantized transport and steady states of Floquet topological insulators,
\href{https://doi.org/10.1103/PhysRevB.97.245401}{Phys. Rev. B \textbf{97}, 245401 (2018)}.
\bibitem{F2}
O. Matsyshyn, F. Piazza, R. Moessner, and I. Sodemann,
Rabi Regime of Current Rectification in Solids,
\href{https://doi.org/10.1103/PhysRevLett.127.126604}{Phys. Rev. Lett. \textbf{127}, 126604 (2021)}.
\bibitem{F3}
T. Morimoto and N. Nagaosa,
Topological nature of nonlinear optical effects in solids,
\href{https://doi.org/10.1126/sciadv.1501524}{Sci. Adv. \textbf{2}, e1501524 (2016)}.
\end{thebibliography}
\end{document}